\documentclass[12pt]{article}
\usepackage[dvips]{epsfig}
\usepackage{graphicx}
\usepackage{latexsym,amsmath,amsfonts,amsthm,amssymb}

\usepackage{graphics}
\usepackage[usenames]{color}

\theoremstyle{plain}
\newtheorem{theorem}{Theorem}[section]

\newcommand{\fz}{{\rm{z}}}

\newcommand{\rarrow}{\rightarrow}

\renewcommand{\theequation}{\thesection.\arabic{equation}}

\title{Dimensional Reduction Formulas for Branched Polymer Correlation Functions}

\author{\begin{tabular}{ccc}
David C. Brydges\thanks{Research supported by NSF Grant DMS-9706166}          & John Z. Imbrie \\
        Department of Mathematics & Department of Mathematics \\
        The University of British Columbia &University of Virginia\\
        Room 121, 1984 Mathematics Road& Charlottesville VA 22904-4137\\
        Vancouver, B.C., Canada V6T 1Z2& {\tt ji2k@virginia.edu}\\
        {\tt db5d@math.ubc.ca}  & \\
        and&\\
Department of Mathematics&\\
       University of Virginia&\\
       Charlottesville VA 22904-4137&\\
\end{tabular}}
       
\date{}

\begin{document}
\maketitle
\thispagestyle{empty}
%\begin{spacing}{2}
\begin{abstract}
In \cite{BI01} we have proven that the generating function for self-avoiding branched polymers in $D+2$ continuum dimensions is proportional to the pressure of the hard-core continuum gas at negative activity in $D$ dimensions. This result explains why the critical behavior of branched polymers should be the same as that of the $i \varphi^3$ (or Yang-Lee edge) field theory in two fewer dimensions (as proposed by Parisi and Sourlas in 1981).

In this article we review and generalize the results of \cite{BI01}. We show that the generating functions for several branched polymers are proportional to correlation functions of the hard-core gas. We derive Ward identities for certain branched polymer correlations. We give reduction formulae for multi-species branched polymers and the corresponding repulsive gases. Finally, we derive the massive scaling limit for the 2-point function of the one-dimensional hard-core gas, and thereby obtain the scaling form of the 2-point function for branched polymers in three dimensions.
\end{abstract}


\section{Introduction and Main Results}\label{s1}
\setcounter{equation}{0}

We define the generating function for branched polymers (mod translations) to be
\begin{equation}\label{eqn1.1}
Z_{\rm{BP}}(\fz) = \sum^\infty_{N=1} \frac{\fz^N}{N!} \sum_T \int_{\mathbb{R}^{(D+2)(N-1)}} dy_2 \ldots dy_N \prod_{ij\in T} [2U'(|y_i-y_j|^2)] \prod_{ij \notin T} U(|y_i-y_j|^2).
\end{equation}
Here $y_1 = 0$, $y_2,\ldots,y_N$ are the positions of the monomers, and we sum over all tree graphs $T$ on $\{1,\ldots,N\}$. We assume that $U(t)$ is a positive weight function which tends to 1 as $t \rarrow \infty$, and that $U'(|y|^2)$ is a positive, integrable function of $y \in \mathbb{R}^{D+2}$. By taking limits, we may take $U(t) = \theta(t-1)$, where $\theta$ is the Heaviside step function. In this case, $2U'(|y_i-y_j|^2) = \delta (|y_i-y_j|-1)$, and we obtain our standard model of hard spheres such that spheres $i$ and $j$ are required to touch if $ij \in T$.

The above definition is a direct translation to the continuum of the familiar model of lattice branched polymers. On the lattice $\mathbb{Z}^{D+2}$, a branched polymer is a finite connected set of nearest-neighbor bonds with no cycles \cite{Sla99}. An $N$-vertex branched polymer is a subset $\{y_1,\ldots,y_N\}$ of $\mathbb{Z}^{D+2}$, together
with a tree graph on $\{y_1,\ldots,y_N\}$ such that for every $\{y_i,y_j\} \in T$, $|y_i-y_j| =1$. One defines $c_N$ to be the number of $N$-vertex branched polymers mod translations. Then, as in \cite{F86}, the generating function $Z_{\rm{BP}}(\fz) = \sum_N \fz^N c_N$ can be written as
\begin{equation}\label{eqn1.2}
Z_{\rm{BP}}(\fz) = \sum^\infty_{N=1} \frac{\fz^N}{N!} \sum_T \sum_{y_{2},\ldots,y_{N}} \prod_{ij\in T} [2U'_{ij}]  \prod_{ij \notin T} U_{ij},
\end{equation}
where $2U'_{ij} = \delta_{|y_{i}-y_{j}|,1}$ and $U_{ij} = 1-\delta_{y_{i},y_{j}}$ enforce the adjacency and loop-free conditions, respectively. For example, $c_3 = 6$ in $\mathbb{Z}^2$, which is correctly accounted for in (\ref{eqn1.2}) as there are 3 trees, 4 possibilities for $y_2$, and then 3 for $y_3$.  For more details, see \cite{BI01}.

Returning to the continuum, we define the partition function for the repulsive gas in a box $\Lambda \subset \mathbb{R}^D$:
\begin{equation}\label{eqn1.3}
Z_{\rm{HC}}(\fz) = \sum^\infty_{N=0} \frac{\fz^N}{N!} \int_{\Lambda^N} dx_1\ldots dx_N 
\prod_{1 \leq i < j \leq N} U(|x_i-x_j|^2).
\end{equation}

The main result of \cite{BI01} is that the identity
\begin{equation}\label{eqn1.4}
\lim_{\Lambda \nearrow \mathbb{R}^D} \frac{1}{|\Lambda|} \log Z_{\rm{HC}} (\fz) = -2 \pi Z_{\rm{BP}} \left( - \, \frac{\fz}{2\pi}\right)
\end{equation}
holds for all $\fz$ such that the right-hand side converges absolutely. The left-hand side of (\ref{eqn1.4}) is $1/(kT)$ times the pressure of the repulsive gas. Evidently, its leading singularity $\sim (\fz-\fz_c)^{2-\alpha_{\rm{HC}}}$ is identical to the leading singularity $\sim(\fz-\fz_c)^{2-\gamma_{\rm{BP}}}$ of $Z_{\rm{BP}}$, where $\fz_c$ is the closest singularity to the origin. Hence
\begin{equation}\label{eqn1.5}
\alpha_{\rm{HC}}(D) = \gamma_{\rm{BP}}(D+2).
\end{equation}
If one can define $\theta$ from the asymptotic form $c_N \sim \fz_c^{-N} N^{-\theta}$, then $\theta = 3-\gamma_{\rm{HC}}$ by an Abelian theorem. Furthermore, one expects that $\sigma$, the Yang-Lee edge exponent, is equal to $1-\alpha_{\rm{HC}}$ \cite[eqn.~7]{PF99}, which with (\ref{eqn1.5}) leads to the Parisi-Sourlas relation \cite{PS81}
\begin{equation}\label{eqn1.6}
\theta(D+2) = \sigma(D)+2.
\end{equation}
One can also see that the exponents $\nu_{\rm{BP}}$, $\eta_{\rm{BP}}$ are equal to their hard-core counterparts in two fewer dimensions (see Section~\ref{s3}).

If one takes
\begin{equation}\label{eqn1.7}
U(|x_i-x_j|^2) = e^{-w(x_i-x_j)},
\end{equation}
with $\hat{w}(k) > 0$, then by the sine-Gordon transformation, (\ref{eqn1.3}) can be written as
\begin{equation}\label{eqn1.8}
Z_{\rm{HC}}(\fz) =\int \exp\left(\int_\Lambda dx \hat{\fz} e^{i\varphi(x)}\right) d \mu_w(\varphi),
\end{equation}
where $d\mu_w$ is the Gaussian measure with covariance $w$, and $\hat{\fz} = \fz e^{w(0)/2}$. Thus certain branched polymer models can be written as $-\hat{\fz} e^{i\varphi}$ field theories in two fewer dimensions. Taking into account an effective mass term $\sim \varphi(x)^2$ from $d\mu_w$, one finds a critical value for $\varphi$ on the imaginary axis, and at the critical $\fz$ the interaction is $i \varphi^3+$ higher order. Thus one expects that these theories are in the same universality class as the Yang-Lee edge $(i\varphi^3)$ theory \cite{Fis78} (see \cite{LF95,PF99} for a more complete investigation of the hypothesis that the repulsive-core singularity is in the Yang-Lee class).
{We note that Shapir \cite{Sha83,Sha85} has given a field theory representation for lattice branched polymers which reduces to the supersymmetric Yang-Lee model of \cite{PS81} when presumably irrelevant terms are dropped.}

Cardy has argued recently \cite{C01} (see also his contribution to this issue) that the crossover from area-weighted self-avoiding loops to ordinary self-avoiding loops in two dimensions is governed by a scaling function related to the Airy function. Part of his argument is the reduction of two-dimensional branched polymers to zero-dimensional $i \varphi^3$ theory. This is, in essence, the content of equations (\ref{eqn1.4}) and (\ref{eqn1.8}).
\bigskip

{\bf Correlation Functions} 

We define first the basic $n$-point density correlations for branched polymers and for repulsive gases. Let
\begin{equation}\label{eqn1.9}
\rho(\tilde{x}) = \sum^N_{i=1} \delta(\tilde{x} - x_i), \qquad 
\rho(\tilde{y}) = \sum^N_{i=1} \delta(\tilde{y} - y_i), 
\end{equation}
where $\tilde{x}, x_i \in \mathbb{R}^D$ and $\tilde{y}, y_i \in \mathbb{R}^{D+2}$. Then we put
\begin{eqnarray}\label{eqn1.10}
G^{(n)}_{\rm{BP}} (\tilde{y}_1,\ldots,\tilde{y}_n;\fz)
& = & \sum^\infty_{N=1} \ \frac{\fz^N}{N!} \sum_T \int_{\mathbb{R}^{(D+2)N}} dy_1 \ldots dy_N
\prod^n_{i=1} \rho(\tilde{y}_i) \prod_{ij \in T} [2U'_{ij}] \prod_{ij \notin T} U_{ij}
\nonumber \\[4mm]
G^{(n)}_{\rm{HC}} (\tilde{x}_1,\ldots,\tilde{x}_n;\fz)
& = & \lim_{\Lambda \nearrow \mathbb{R}^{D}} \left\langle 
\prod^n_{i=1} \rho(\tilde{x}_i)\right\rangle_{\rm{HC},\Lambda}.
\end{eqnarray}
Here $U'_{ij} := U' (|y_i-y_j|^2)$, $U_{ij} := U(|y_i-y_j|^2)$, and $\langle \cdot \rangle_{\rm{HC},\Lambda}$ is the expectation in the measure for which $Z_{\rm{HC}}(\fz)$ is the normalizing constant. We also write $G_{\rm{HC}}^{(n),{\rm{T}}}$ for the corresponding truncated expectation.

If $y_1,\ldots,y_n$ are distinct points, then we have
\begin{equation}\label{eqn1.11}
G^{(n)}_{\rm{BP}} (y_1,\ldots,y_n;\fz)
 =  \sum^\infty_{M=0} \ \frac{\fz^M}{M!} \sum_{T \, {\rm on} \, \{1,\ldots,n+M\}} \int dy_{n+1} \ldots dy_{n+M}
\prod_{ij\in T} [2U'_{ij}] \prod_{ij \notin T} U_{ij}.
\end{equation}
Thus, for distinct points $G_{\rm{BP}}^{(n)}$ is a sum/integral over branched polymers whose vertices include $y_1,\ldots,y_n$. 
When points are not distinct, $G_{\rm{BP}}^{(n)}$ and $G_{\rm{HC}}^{(n)}$ are understood by smearing each $\rho(\tilde{y})$ or $\rho(\tilde{x})$ by test functions.  Thus in general, $G_{\rm{BP}}^{(n)}$ and $G_{\rm{HC}}^{(n)}$ are distributions which contain $\delta$-function singularities at coinciding points. In addition, if $U'_{ij}$ is not smooth (for example in the hard sphere case $U(t) = \theta(t-1)$), then $G_{\rm{BP}}^{(n)}$ will inherit singularities from $U'_{ij}$.

The density correlations $G^{(n)}_{\rm{HC}}$ arise naturally when taking an order $n$ variational derivative of $Z_{\rm{HC}}$ with respect to an external field.  However, we will need a different set of Green's functions for the repulsive gas. Stripping $G_{\rm{HC}}^{(n)}$ of its singularities at coinciding points, we write
\begin{equation}\label{eqn1.12}
g^{(n)}_{\rm{HC}} (x_1,\ldots,x_n;\fz)
 = Z_{\rm{HC}}(\fz)^{-1} \sum^\infty_{m=0} \ \frac{\fz^{n+m}}{m!}  \int_{\Lambda^m}  dx_{n+1} \ldots dx_{n+m}
\prod_{1\leq i < j \leq n+m} U_{ij}.
\end{equation}
Here $n$ particles are forced to be at $x_1,\ldots,x_n$, and if these are distinct points, then $g_{\rm{HC}}^{(n)} = G_{\rm{HC}}^{(n)}$. In general, $G_{\rm{HC}}^{(n)}$ is a sum of terms, each with some $g_{\rm{HC}}^{(j)}$, $j \leq n$ multiplied by $\delta$-functions in some of the $x_i$'s. For example,
\begin{equation}\label{eqn1.13}
G_{\rm{HC}}^{(2)}(x_1,x_2) = g_{\rm{HC}}^{(2)} (x_1,x_2) + g^{(1)}_{\rm{HC}}(x_1) \delta(x_1-x_2).
\end{equation}
We shall see that $g_{\rm{HC}}^{(n)}$ can be related to a certain $n$-tree branched polymer correlation function:
\begin{equation}\label{eqn1.14}
g_{\rm{BP}}^{(n)} (y_1,\ldots,y_n;\fz) = \sum^\infty_{p=0} \frac{\fz^{p+n}}{p!} \sum_{F^{(n)}}
\int_{\mathbb{R}^{(D+2)p}} dy_{n+1} \ldots dy_{n+p} \prod_{ij \in F^{(n)}} [2U'_{ij}] \prod_{ij \notin F^{(n)}} U_{ij}.
\end{equation}
Here $F^{(n)}$ is a loop-free graph or {\em forest} on $\{1,\ldots,n+p\}$ which consists of $n$ connected components or {\em trees}, each of which contains one of $y_1,\ldots,y_n$.

\begin{theorem} \label{th1}
If $\fz$ is in the interior of the domain of convergence at $Z_{\rm{BP}}$, then in the limit $\Lambda \nearrow \mathbb{R}^D$,
\begin{equation}\label{eqn1.15}
g_{\rm{HC}}^{(n)} (x_1,\ldots,x_n;\fz) = (-2\pi)^n
g_{\rm{BP}}^{(n)}\bigg(x_1,\ldots,x_n;-\frac{\fz}{2\pi}\bigg),
\end{equation}
and
\begin{equation}\label{eqn1.16}
G_{\rm{HC}}^{(n),{\rm{T}}} (x_1,\ldots,x_n;\fz) = (-2\pi) \int_{\mathbb{C}^{n-1}} dz_2 \ldots dz_n
G_{\rm{BP}}^{(n)}\bigg(x_1,y_2,\ldots,y_n; -\frac{\fz}{2\pi}\bigg).
\end{equation}
Here $x_i \in \mathbb{R}^D$ and $y_i = (x_i,z_i) \in \mathbb{R}^{D+2}$.
\end{theorem}

The relation (\ref{eqn1.16}) was proven in \cite{BI01} by differentiating (\ref{eqn1.4}) with respect to sources. We prove (\ref{eqn1.15}) in Section~\ref{s2}. As $g_{\rm{HC}}^{(n)}$ and $G_{\rm{HC}}^{(n)}$ agree at non-coinciding points, (\ref{eqn1.15}) and (\ref{eqn1.16}) combine to give relations between
$g_{\rm{BP}}^{(n)}$ and $G_{\rm{BP}}^{(n)}$. In particular, we show that the two-point functions obey a Ward identity
\begin{equation}\label{eqn1.17}
\frac{d}{d(r^2)} \ g^{(2)}_{\rm{BP}} = \frac{1}{2} G_{\rm{BP}}^{(2)},
\end{equation}
where by rotation and translation invariance
$g^{(2)}_{\rm{BP}}$ and $G_{\rm{BP}}^{(2)}$ can be thought of as functions of $r^2 = |y_1-y_2|^2$ only.

In an appendix, we generalize the above results to repulsive gases and branched polymers with more than one species of particle/monomer and species-dependent interactions. Examples include the Widom-Rowlinson model of penetrable hard spheres
\cite{WR70}. As with the models discussed above, dimensional reduction is actually a consequence of an underlying supersymmetry of the branched polymer model. This requires that the attractive interaction between neighboring monomers be related to the repulsive interaction.

In Section~\ref{s3} we focus on the case $D=1$ and derive a number of results for the standard hard-core gas using the method of Laplace transforms. We give fairly explicit formulas for $G_{\rm{HC}}^{(2),{\rm{T}}}$ and derive the values $\alpha_{\rm{HC}} = \frac{3}{2}$, $\nu_{\rm{HC}} = \frac{1}{2}$, $\eta_{\rm{HC}} = -1$, thereby obtaining the same values for $\gamma_{\rm{BP}}$, $\nu_{\rm{BP}}$, $\eta_{\rm{BP}}$ in three dimensions. (Note that for $D=0$, $\log Z_{\rm{HC}} = \log(1+z)$, so that the two-dimensional $Z_{\rm{BP}}(\fz)$ has a logarithmic singularity at $\fz = 2\pi$, which implies that $\gamma_{\rm{BP}}=\alpha_{\rm{HC}} = 2$. Unfortunately, dimensional reduction gives no information on $\nu_{\rm{BP}}$, $\eta_{\rm{BP}}$ in this case.)  We derive the scaling form of the two-point function near $\fz_c = -e^{-1}$:
\begin{equation}\label{eqn1.18}
G^{(2),{\rm{T}}}_{\rm{HC}}(0,x;\fz) \sim |x|^{-(D-2+\eta_{\rm{HC}})} K_{\rm{HC}} (x/\xi),
\end{equation}
with
\begin{equation}\label{eqn1.19}
K_{\rm{HC}}(\hat{x}) = -\frac{4}{\hat{x}^2} \ e^{-\hat{x}},
\end{equation}
which implies
\begin{equation}\label{eqn1.20}
G^{(2)}_{\rm{BP}}(0,y;\fz) \sim |x|^{-d-2+\eta_{\rm{BP}}} K_{\rm{BP}} (x/\xi),
\end{equation}
with
\begin{equation}\label{eqn1.21}
K_{\rm{BP}}(\hat{x}) = \frac{1}{\pi^2\hat{x}} \ e^{-\hat{x}}.
\end{equation}
The form of $K_{\rm{HC}}(\hat{x})$ is the same as that of the one-dimensional Ising model near the Yang-Lee edge \cite{Fis80}. The form of $K_{\rm{BP}}(\hat{x})$ agrees with the prediction of Miller \cite{Mil91}.

\section{The Forest-Root Formula and Dimensional Reduction} \label{s2}
\setcounter{equation}{0}

We wish to derive relationships between the hard-core Green's functions in $D$ dimensions and the branched polymer Green's functions in $D+2$ dimensions. The key is the Forest-Root formula, proven in \cite{BI01}. Let $f(\bf{t})$ be any smooth function of variables
$$
t_{ij} = |z_i-z_j|^2, \qquad 1 \leq i < j \leq N
\quad \mbox{ and } \quad t_i = |z_i|^2, \qquad 1 \leq i \leq N,
$$
where each $\fz_i \in \mathbb{C}$. Assume that $f$, when regarded as a function of $\fz_1,\ldots,\fz_N$, has compact support. Any subset of the bonds $\{ij| 1 \leq i < j \leq N\}$ forms a graph on the vertices $\{1,\ldots,N\}$. A subset $R$ of vertices is called a set of {\em roots}. A {\em forest} $F$ is a graph that has no loops. The connected components of a forest are {\em trees}. We are declaring that a graph with no bonds and just one vertex is also a tree. See Fig.~1.

\begin{figure}[h]
\centering
\includegraphics[width=.9\textwidth]{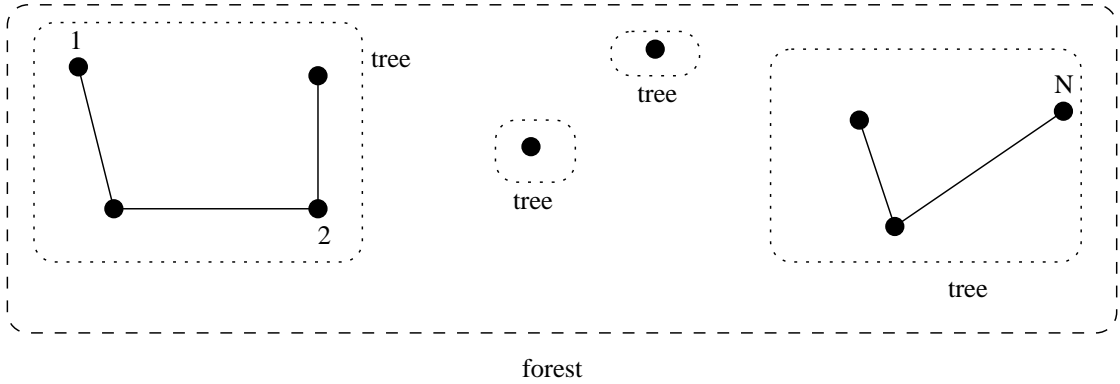}
\caption{Example of a forest}
\end{figure} 

\begin{theorem} \label{th2}(Forest-Root Formula)
\begin{equation}\label{eqn2.1}
f({\bf{0}}) = \sum_{(F,R)} \int_{\mathbb{C}^N} f^{(F,R)} ({\bf{t}})
\left(\frac{d^2z}{-\pi}\right)^N,
\end{equation}
where $f^{(F,R)}({\bf{t}})$ denotes the derivative with respect to the variables $t_{ij}$ with $ij \in F$ and $t_i$ with $i \in R$. The sum is over all forests $F$ and all sets $R$ of roots with the property that each tree in $F$ contains exactly one root from $R$, and $d^2z = du \ dv$, where $z = u+iv$.
\end{theorem}

\emph{Proof of Theorem \ref{th1} (\ref{eqn1.15}).} In order to examine $g^{(n)}_{\rm{HC}}$, we set $N = n+m$ and put
\begin{equation}\label{eqn2.2}
f({\bf{t}}) = \prod_{1 \leq i < j \leq N} U(|x_{ij}|^2+t_{ij}) \prod^m_{i=1} g(\epsilon t_i) \prod^n_{j=1} g(t_j/\epsilon),
\end{equation}
where $x_{ij} = x_i-x_j$, $ \epsilon > 0$, and where $g$ is a smooth, decreasing function with compact support such that $g(0) = 1$. Working in a finite box $\Lambda$, we write
\begin{equation}\label{eqn2.3}
g_{\rm{HC}}^{(n)}(x_1,\ldots,x_n) Z_{\rm{HC}}(\fz) = \sum^\infty_{m=0} \frac{\fz^{m+n}}{m!} \int_{\Lambda^n} dx_{n+1} \ldots dx_{n+m} f({\bf{0}}),
\end{equation}
and insert (\ref{eqn2.1}). With $y_i = (x_i,z_i)$, this becomes
\begin{equation}\label{eqn2.4}
\sum^\infty_{m=0} \frac{\fz^{m+n}}{m!} \int_{\Lambda^n} 
dx_{n+1} \ldots dx_{n+m} 
\sum_{(F,R)} \int_{\mathbb{C}^N}
\left(\frac{d^2z}{-\pi}\right) \prod_{ij \in F} U'_{ij} \prod_{ij \notin F} U_{ij} \prod_{i\in R} \frac{dg}{dt_i} \prod_{i \notin R} g.
\end{equation}
The forest $F$ may be divided into $F^{(n)}$ (which consists of all the trees containing any of the vertices $1,\ldots,n$), and the rest, $\tilde{F}$. If any of the vertices $1,\ldots,n$ are not roots, then the corresponding factor $g(t/\epsilon)$ is not differentiated, and so lacks a factor $\epsilon^{-1}$ to compensate for the $O(\epsilon)$ integration volume. Therefore, neglecting $O(\epsilon)$ terms, $F^{(n)} = T_1 \cup \cdots\cup T_n$ with $x_i \in T_i$ for $i = 1,\ldots,n$ , and $T_{i}$ disjoint. Let $m = p+M$, where $M$ is the number of vertices in $\tilde{F}$, and rewrite (\ref{eqn2.4}) as
\begin{eqnarray}\label{eqn2.5}
\lefteqn{\hspace*{-30pt}
\sum^\infty_{p=0} \frac{z^{p+n}}{p!} \int_{\Lambda^p} dx_{n+1} \ldots dx_{n+p} 
\sum_{F^{(n)}} \int_{\mathbb{C}^p} \left(\frac{d^2z}{-\pi}\right)^p \prod_{ij \in F^{(n)}} U'_{ij}
} \nonumber \\[4mm]
& \cdot & \sum^\infty_{M=0} \frac{\fz^{M}}{M!} \int_{\Lambda^M} dx_{n+p+1} \ldots dx_{n+p+M} 
\sum_{(\tilde{F},\tilde{R})} \int_{\mathbb{C}^M} \left(\frac{d^2z}{-\pi}\right)^M \prod_{ij \in \tilde{F}} U'_{ij}
\nonumber \\[4mm]& \cdot & 
\prod_{i \in \tilde{R}} [\epsilon g'(\epsilon t_i)] \prod_{i \notin R} g(\epsilon t_i) \prod_{ij \notin F} U_{ij} + o(1),
\end{eqnarray}
where $o(1)$ denotes a quantity which tends to zero with $\epsilon$.  Here $\tilde{R}$ is a subset of $\{n+p+1,\ldots,n+p+M\}$ and as before, each tree of $\tilde{F}$ has exactly one root from $\tilde{R}$. We have eliminated the integrals over $z_1,\ldots,z_n$ because ${(-\pi \epsilon)^{-1} g'(t_i/\epsilon)}$ tends to $\delta(z_i)$. The factors $g(\epsilon t_i)$ with $n + 1 \leq i \leq n+p$ can be replaced with 1 because the decrease of $U'$ in essence forces the corresponding $z_i$'s to remain bounded. Any errors from these approximations are $o(1)$.

The only barrier to writing (\ref{eqn2.5}) as a product is the presence of interactions $U_{ij}$ linking $F^{(n)}$ and $\tilde{F}$. However, for small $\epsilon$ the trees in $F$ are rarely close to each other or to the trees of $F^{(n)}$. Using again the decrease of $U'$ we see that all the vertices of a tree are in a bounded cluster and that $g(\epsilon t_i)$ can be replaced with $g(\epsilon t_r)$, where $r$ is the root of the tree. Then the sum over roots leads to a factor $N(T)$, the number of vertices in $T$. Observe that $(-\epsilon/\pi)N(T) g(\epsilon t_r)^{N(T)-1}g'(\epsilon t_r) d^2 z_r$ is a probability measure which becomes very wide as $\epsilon \rarrow 0$. Thus with high probability, the interactions $U_{ij}$ between $F^{(n)}$ and $\tilde{F}$ can be replaced with 1, with an additional contribution to the $o(1)$ error. As a result, (\ref{eqn2.5}) can be rewritten as
\begin{equation}\label{eqn2.6}
\left[
\sum^\infty_{p=0} \frac{\fz^{p+n}}{p!} \int_{(\Lambda\times \mathbb{C})^p} 
\prod^{n+p}_{i=n+1} \frac{d^{D+2}y_i}{-\pi} \sum_{F^{(n)}} \prod_{ij \in F^{(n)}} U'_{ij}
\prod_{ij \notin F^{(n)}} U_{ij}\right] Z_{\rm{HC}} (\fz) +o(1).
\end{equation}
Taking the limit as $\epsilon \rarrow 0$, we obtain a relation for finite $D$-dimensional volume $\Lambda$:
\begin{equation}\label{eqn2.7}
g_{\rm{HC}}^{(n)} (x_1,\ldots,x_n;\fz) = g_{\rm{BP}}^{(n)} \left(x_1,\ldots,x_n;-\frac{\fz}{2\pi}\right)(-2\pi)^n.
\end{equation}
The limit $\Lambda \nearrow \mathbb{R}^D$ exists for each term in the sum over $p$, by monotone convergence. By dominated convergence, the sum on $p$ may be interchanged with the infinite volume limit, and we obtain the first part of Theorem \ref{th1}.

\emph{Proof of (\ref{eqn1.4}).} A similar factorization occurs in $Z_{\rm{HC}}(\fz)$, so that all the terms with $k$ trees can be written as $1/k!$ times the $k^{\text{th}}$ power of
\begin{equation}\label{eqn2.8}
\sum^\infty_{N=1} \frac{\fz^N}{N!} \sum_T \int_\Lambda dx_1 \int_{(\Lambda \times \mathbb{C})^{N-1}} \prod^N_{i=2} \frac{dy_i}{-\pi}
\prod_{ij \in T} U'_{ij} \prod_{ij \notin T} U_{ij},
\end{equation}
so that, as argued in \cite{BI01},
\begin{equation}\label{eqn2.9}
\lim_{\Lambda \nearrow \mathbb{R}^D} \frac{1}{|\Lambda|} \log Z_{\rm{HC}} (\fz) = -2\pi Z_{\rm{BP}} \left( -\frac{\fz}{2\pi}\right).
\end{equation}

\emph{Proof of Theorem~\ref{th1} (\ref{eqn1.16}).} The relation between $G_{\rm{HC}}^{(n)}$ and $G_{\rm{BP}}^{(n)}$ may be derived by differentiating (\ref{eqn2.9}) with respect to sources. Then, as explained in \cite{BI01},
\begin{equation}\label{eqn2.10}
G_{\rm{HC}}^{(n)} (x_1\ldots x_n;\fz) = (-2\pi) \int_{\mathbb{C}^{n-1}} \prod^n_{i=2} d^2z_i \   G_{\rm{BP}}^{(n)}\left(x_1,y_2,\ldots,y_n;-\frac{\fz}{2\pi}\right),
\end{equation}
which is (\ref{eqn1.16}). In momentum space, then, $G_{\rm{HC}}^{(n),{\rm{T}}}$ may be obtained from $G_{\rm{BP}}^{(n)}$ by setting the components of momenta in the two extra dimensions to zero. This contrasts with the relation (\ref{eqn2.7}) between $g_{\rm{HC}}^{(n)}$ and $g_{\rm{BP}}^{(n)}$, in which the spatial components in the two extra dimensions are set to 0.

\emph{Proof of (\ref{eqn1.17}).} Relations between $G_{\rm{BP}}$ and $g_{\rm{BP}}$ may be derived by combining (\ref{eqn2.7}) and (\ref{eqn2.10}). For example, consider the 2-point functions, which by rotation invariance can be expressed as functions of the squared-distance $t$:
\begin{eqnarray}\label{eqn2.11}
G_{\rm{HC}}^{(2)}(t;\fz) & := & G_{\rm{HC}}^{(2)} (x_1,x_2;\fz),
\mbox{ where } |x_1-x_2|^2=t,
\nonumber \\[3mm]
G_{\rm{BP}}^{(2)}(t;\fz) & := & G_{\rm{BP}}^{(2)} (y_1,y_2;\fz),
\mbox{ where } |y_1-y_2|^2=t,
\end{eqnarray}
and similarly for $g_{\rm{HC}}^{(2)}$ and $g_{\rm{BP}}^{(2)}$. Since $g_{\rm{HC}}^{(2)}$ agrees with $G_{\rm{HC}}^{(2)}$ at non-coinciding points, (\ref{eqn2.7}) and (\ref{eqn2.10}) imply that
\begin{equation}\label{eqn2.12}
g_{\rm{BP}}^{(2),{\rm{T}}}(t;\fz) = (-2\pi)^{-1} \int^\infty_t \pi dt' \ 
G_{\rm{BP}}^{(2)}(t',\fz), \quad t \neq 0.
\end{equation}
Differentiation yields
\begin{equation}\label{eqn2.13}
\frac{d}{dt} g_{\rm{BP}}^{(2)}(t;\fz) = \frac{1}{2} 
G_{\rm{BP}}^{(2)}(t;\fz).
\end{equation}
This may be thought of as a Ward identity for the supersymmetry of our model of branched polymers.

\section{Green's Function for the Hard-Core Gas in One Dimension}\label{s3}
\setcounter{equation}{0}

Laplace transforms can be used to give fairly explicit formulas for the Green's function for one-dimensional gases with only nearest neighbor interactions. We follow \cite{FW69} in deriving the relevant expressions for the basic hard-core gas with no interactions other than a minimum separation of 1 between particles.

Let us write the grand canonical partition function in the following way (we omit the subscript $\rm{HC}$ in most of this section):
\begin{equation}\label{eqn3.1}
Z(L) = \sum^\infty_{N=0} \fz^N \int_{x_{1}\geq 1} dx_1 \int_{x_{2} \geq x_{1}+1} dx_2 \cdots \int_{L-1 \geq x_{N} \geq x_{N-1} +1} dx_N.
\end{equation}
The particles are restricted to the interval $\Lambda = [1,L-1]$, as if external particles had been placed at 0 and $L$. We assume $L > 1$ and put $Z(L) = 1$ for $1 < L \leq 2$. The Laplace transform can be evaluated explicitly:
\begin{eqnarray}\label{eqn3.2}
\hat{Z}(s) & := & \int^\infty_1 dL \, e^{-sL} Z(L)
\nonumber \\[4mm]
& = & \sum^\infty_{N=0} \fz^N J(s)^{N+1},
\end{eqnarray}
where
\begin{equation}\label{eqn3.3}
J(s) = \int^\infty_1 dx \, e^{-sx} = \frac{1}{s} \, e^{-s}.
\end{equation}
Using analytic continuation as necessary to define $\hat{Z}(s)$, we have
\begin{equation}\label{eqn3.4}
\hat{Z}(s) = \frac{J(s)}{1-\fz J(s)} = \frac{1}{se^s-\fz}.
\end{equation}
We obtain $Z(L)$ by inverse transform:
\begin{equation}\label{eqn3.5}
Z(L) = \frac{1}{2\pi i} \int \frac{1}{se^s-\fz} \ e^{sL} ds.
\end{equation}
This leads to a residue formula
\begin{equation}\label{eqn3.6}
Z(L) = \sum^\infty_{n=0} \frac{e^{s_n(L-1)}}{s_n+1},
\end{equation}
where $\{s_n\}$ are the solutions to $se^s=\fz$, arranged in order of decreasing real part.
These solutions are the branches of the Lambert $W$-function \cite{CGHJK}.

We will make use of some properties of the $s_n$. For $\fz > 0$, there is one real solution, and for $-e^{-1} < \fz < 0$, there are two real solutions (see Fig.~2). The complex solutions come in conjugate pairs, and all have real parts which are less than the real solutions. (This can be seen by writing $s = x+iy$ and letting $x(y)$ solve the modulus equation $(x^2+y^2) e^{2x} = |\fz|^2$. Then $\frac{dx}{dy^2} = -1/[2(x+x^2+y^2)] < -c < 0$ with $c$ independent of $x,y$ in any bounded region not intersecting $\{(x,y)|x \in (-1,0)\}$. This shows, in fact, that the upper gap $\mbox{Re}(s_1-s_2) > B > 0$ with $B$ independent of $\fz$ in any interval $[-1,\fz_0]$ with $\fz_0 < 0$.) If we put $s_n = x_n +iy_n$, then $|y_n-n\pi| \leq $ const \cite[Fig.~4]{CGHJK}. In addition, $x_n \sim -\log |y_n/\fz| \sim -\log|n/\fz|$ for large $n$, from the modulus equation. Hence the sum in (\ref{eqn3.6}) converges for all $L > 1$.

\begin{figure}
\begin{center}
\begin{picture}(160,160)
\put(0,75){\line(1,0){150}}
\put(75,0){\line(0,1){150}}
\multiput(40,75)(0,-9){4}{\line(0,-1){5}}
\multiput(40,40)(9,0){4}{\line(1,0){5}}
\bezier{1000}(25,50)(10,70)(0,70)
\bezier{1000}(75,75)(42,23)(25,50)
\bezier{1000}(105,155)(95,115)(75,75)
\put(77,148){$\fz$}
\put(36,78){$-1$}
\put(77,40){$-e^{-1}$}
\put(146,68){$s$}
\end{picture}
\caption{Graph of the function $\fz = se^s$}
\end{center}
\end{figure}
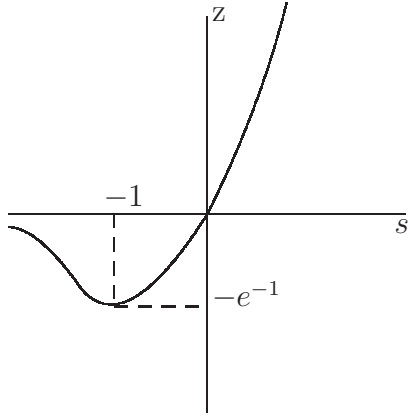

The density, or one-point function, $G^{(1)}(x) = G^{(1)}$, is the expectation of $\rho(x) = \sum^N_{j=1} \delta(x-x_j)$ in the limit as $\Lambda \nearrow \infty$. If we take $\Lambda = \left[- \frac{L}{2} +1, \frac{L}{2}-1\right]$, then
\begin{equation}\label{eqn3.7}
G^{(1)} = \lim_{L \rarrow \infty} \frac{Z\left(\frac{L}{2}+x\right)\fz Z\left(\frac{L}{2}-x\right)}{Z(L)}.
\end{equation}
Only the $n=0$ term of (\ref{eqn3.6}) survives the $L \rarrow \infty$ limit. Thus we have
\begin{equation}\label{eqn3.8}
G^{(1)} = \frac{\fz e^{-s_0}}{s_0+1} = \frac{s_0}{s_0+1},
\end{equation}
where we have used the relation $\fz = s_0 e^{s_0}$.  If we identify $G^{(1)}$ with the density $\bar{\rho}$ and solve for the pressure $p=s_0kT$, we obtain the equation of state for hard rods of unit length \cite{Ton36}:
\begin{equation}
p=\frac{\bar{\rho} kT}{1-\bar{\rho}}.
\end{equation}

The density-density correlation, or two-point function $G^{(2)}(0,x) = G^{(2)}(x)$, is the expectation of $\rho(0) \rho(x)$. For $x > 1$, this can be written as
\begin{eqnarray}\label{eqn3.9}
G^{(2)}(x)
& = & \lim_{L\rarrow \infty}
\frac{Z\left(\frac{L}{2}\right)\fz Z(x)\fz Z\left(\frac{L}{2}-x\right)}{Z(L)}
\nonumber \\[4mm]
& = & \frac{s_0}{s_0+1} e^{-s_0x} \fz Z(x).
\end{eqnarray}

One can insert the formula
\begin{equation}\label{eqn3.10}
Z(x) = \sum^\infty_{N=0} \frac{\fz^N}{N!} (x-N-1)^N \theta(x-N-1)
\end{equation}
to obtain the long-known expression for $G^{(2)}$ (see, for example, \cite[eqn.~32]{SZK53}) which is useful if $x$ is not too large. Alternatively, one can insert (\ref{eqn3.6}) to obtain
\begin{equation}\label{eqn3.11}
G^{(2)}(x) = \sum^\infty_{n=0} \frac{s_0}{s_0+1} \ \frac{s_n}{s_n+1} \ e^{(s_n-s_0)x},
\end{equation}
and after subtracting $G^{(1)2}$, we obtain an expression for the truncated Green's function
\begin{equation}\label{eqn3.12}
G^{(2),{\rm{T}}}(x) = \sum^\infty_{n=1} \frac{s_0}{s_0+1} \ \frac{s_n}{s_n+1} \ e^{(s_n-s_0)x},
\end{equation}
which is convergent for $x > 1$. It is apparent from (\ref{eqn3.10}) that $\left(\frac{d}{dx}\right)^N G^{(2),{\rm{T}}}(x)$ is continuous, except for a jump at $x = N+1$. This is reflected in the divergence of the series (\ref{eqn3.12}) at $x = N+1$, when differentiated $N$ times. For $|x| < 1$, $G^{(2),{\rm{T}}}(x) = 0$ except for a $\delta$-function at 0 with coefficient $G^{(1)}$.
\bigskip

{\bf Scaling Form of the Green's Function}

As $\fz \searrow \fz_c = -e^{-1}$, the two real solutions to $se^s=\fz$ approach the value $-1$. The correlation length is given by
\begin{equation}\label{eqn3.13}
\xi := \left[\lim_{|x|\rarrow \infty} - \frac{1}{|x|} \log G^{(2),{\rm{T}}}(x)\right]^{-1} = (s_0-s_1)^{-1}.
\end{equation}
It is clear from Fig.~2 that $s_0-s_1 \sim (\fz-\fz_c)^{\frac{1}{2}}$. Hence, as $\fz \searrow \fz_c$, $\xi$ diverges as $(\fz-\fz_c)^{-\nu}$ with a correlation exponent $\nu = \frac{1}{2}$. 

If we let $\fz \searrow \fz_c$ and $x \rarrow \infty$ while keeping $\hat{x} = x/\xi$ fixed, then the asymptotic form of $G^{(2),{\rm{T}}}$ is described by a scaling function
\begin{equation}\label{eqn3.14}
K(\hat{x}) = \lim_{x \rarrow \infty, \fz \searrow \fz_c}
x^{D-2+\eta} G^{(2),{\rm{T}}}(x).
\end{equation}
Here $D=1$, and we take the anomalous dimension $\eta =-1$ in order to get a nontrivial limit. From (\ref{eqn3.12}), and the uniform gap between $s_1$ and the other solutions, we have
\begin{equation}\label{eqn3.15}
K(\hat{x}) = \lim_{x \rarrow \infty, \fz \searrow \fz_c}
\frac{s_0}{s_0+1} \ \frac{s_1}{s_1+1} \ \frac{1}{x^2} \ e^{-x/\xi}.
\end{equation}
A short calculation shows that
$\frac{s_0}{s_0+1} \ \frac{s_1}{s_1+1} = -\frac{4}{(s_0-s_1)^2}
(1+O(\fz-\fz_c))$, and hence
\begin{equation}\label{eqn3.16}
K(\hat{x}) =  -\frac{4}{\hat{x}^2} \ e^{-\hat{x}}.
\end{equation}

We may also define a scaling function for branched polymers:
\begin{equation}\label{eqn3.17}
K_{\rm{BP}}(\hat{x}) = \lim_{x \rarrow \infty, \fz \nearrow \tilde{\fz}_c} x^{d-2+\eta_{\rm{BP}}} G_{\rm{BP}}^{(2)} (0,x).
\end{equation}
Here $\hat{x} = x/\xi_{\rm{BP}}$ with
\begin{equation}\label{eqn3.18}
\xi_{\rm{BP}} := \left[\lim_{|x| \rarrow \infty} - \frac{1}{|x|}
\log G_{\rm{BP}}^{(2)} (0,x)\right]^{-1},
\end{equation}
and $\eta_{\rm{BP}}$ is chosen so as to obtain a nontrivial limit for $K_{\rm{BP}}$. As explained in \cite{BI01}, we have the relation (\ref{eqn2.10}) between $G^{(2)}_{\rm{BP}}$ and $G^{(2),{\rm{T}}}_{\rm{HC}}$, which when differentiated yields
\begin{equation}\label{eqn3.19}
G_{\rm{BP}}^{(2)} \left(t; -\frac{\fz}{2\pi}\right) = \frac{1}{2\pi^2} \frac{d}{dt} \, G_{\rm{HC}}^{(2),{\rm{T}}}(t,\fz).
\end{equation}
Hence, the critical activity $\tilde{\fz}_c$ for branched polymers is equal to $-2\pi \fz_c = 2\pi e$, $\xi_{\rm{BP}}\left(-\frac{\fz}{2\pi}\right) = \xi_{\rm{HC}}(\fz), \eta_{\rm{BP}} = \eta_{\rm{HC}}$, and
\begin{eqnarray}\label{eqn3.20}
K_{\rm{BP}}(\hat{x})
& = & \frac{1}{4\pi^2} [\hat{x} K'_{\rm{HC}}(\hat{x})-(D-2+\eta_{\rm{HC}}) K_{\rm{HC}}(\hat{x})]
\nonumber \\[4mm]
& = & \frac{1}{\pi^2 \hat{x}} \ e^{-\hat{x}}.
\end{eqnarray}
Of course, the exponent $\nu_{\rm{BP}}$ governing the divergence of $\xi_{\rm{BP}}$ as $\fz \nearrow \tilde{\fz}_c$ must equal $\nu_{\rm{HC}}$ ($= \frac{1}{2}$ if $d = D+2=3$).
\bigskip

\section*{Appendix: Multispecies Examples}\label{sa}
\renewcommand{\theequation}{{\rm A}.\arabic{equation}}
\setcounter{equation}{0}

We can generate the arguments of Section~\ref{s2} to multispecies examples. Define a repulsive gas partition function in $D$ dimensions:
\begin{equation}\label{eqnA.1}
Z_{\rm{HC}}({\bf z}) = \sum^\infty_{N=0} \ \frac{1}{N!} \ \sum_{\alpha_{1},\ldots,\alpha_{N}} \prod^N_{i=1} \fz_\alpha \int_{\Lambda^N} dx_1 \cdots dx_N \prod_{ij} U^{\alpha_i \alpha_j} (|x_i-x_j|^2),
\end{equation}
where each $\alpha_i$ is summed over the set of species of the problem, $\fz_\alpha$ is the activity of species $\alpha$, and $U^{\alpha\beta}$ is a repulsive interaction between species $\alpha$ and species $\beta$.
The corresponding multispecies branched polymer generating function is
\begin{equation}\label{eqnA.2}
Z_{\rm{BP}}({\bf z}) = \sum^\infty_{N=1} \ \frac{1}{N!} \ \sum_T \ \sum_{\alpha_{1},\ldots,\alpha_{N}} \prod^N_{i=1} \fz_\alpha \int_{(\mathbb{R}^{D+2})^{N-1}} dy_2 \cdots dy_N \prod_{ij \in T} [2U'_{ij}] \prod_{ij \notin T} U_{ij},
\end{equation}
where
\begin{equation}\label{eqnA.3}
U_{ij} = U^{\alpha_i \alpha_j} (|x_i-x_j|^2),
\end{equation}
and $U'_{ij}$ is its derivative. In particular, for the hard-core model with minimum separation $R_{\alpha\beta}$ between species $\alpha$ and species $\beta$, we would have $U^{\alpha\beta}(t) = \theta(t-R^2_{\alpha\beta})$ and
\begin{equation}\label{eqnA.4}
2 U^{\alpha \beta}{'}(|x_i-x_j|^2) = \frac{1}{R_{\alpha\beta}} \delta(|x_i-x_j|-R_{\alpha \beta}).
\end{equation}

Assume that $U^{\alpha \beta}$ satisfies the usual conditions ($U^{\alpha \beta}, U^{\alpha \beta}{'}$ positive, $U^{\alpha \beta} \rarrow 1$ at $\infty$, $U^{\alpha\beta}{'}$ integrable in $\mathbb{R}^{D+2}$). Then, provided $Z_{\rm{BP}}({\bf z})$ is absolutely convergent, we obtain a reduction formula
\begin{equation}\label{eqnA.5}
\lim_{\Lambda \nearrow \mathbb{R}^D} \frac{1}{|\Lambda|} \log Z_{\rm{HC}}({\bf z}) = -2\pi Z_{\rm{BP}} \left(- \frac{\bf z}{2\pi}\right).
\end{equation}
We also obtain results as in Theorem \ref{th1} for correlation functions.

{\bf Acknowledgement}

We thank Yonathan Shapir and Joel Lebowitz for useful suggestions which improved the paper.

\newcommand{\etalchar}[1]{$^{#1}$}

%\end{spacing}


\begin{thebibliography}{CGHJK}

\bibitem[BI01]{BI01}
D.C. Brydges and J.Z. Imbrie.
\newblock Branched polymers and dimensional reduction.
\newblock Preprint, arXiv:math-ph/0107005.


\bibitem[Car01]{C01}
J.L. Cardy.
\newblock Exact scaling functions for self-avoiding loops and branched
  polymers.
\newblock {\em J. Phys. A}, 34:L665--L672, 2001.
\newblock arXiv:cond-mat/0107223.

\bibitem[CGHJK]{CGHJK}
R.~M. Corless, G.~H. Gonnet, D.~E.~G. Hare, D.~J. Jeffrey, and D.~E. Knuth.
\newblock On the {L}ambert ${W}$ function.
\newblock {\em Adv. Comput. Math.}, 5:329--359, 1996.

\bibitem[Fis78]{Fis78}
M.E. Fisher.
\newblock {Y}ang-{L}ee edge singularity and $\varphi^3$ field theory.
\newblock {\em Phys. Rev. Lett.}, 40:1610--1613, 1978.

\bibitem[Fis80]{Fis80}
M.E. Fisher.
\newblock {Y}ang-{L}ee edge behavior in one-dimensional systems.
\newblock {\em Prog. Theor. Phys. Suppl.}, 69:14--29, 1980.

\bibitem[Fr{\"o}86]{F86}
J.~Fr{\"o}hlich.
\newblock Mathematical aspects of the physics of disordered systems.
\newblock In {\em Ph\'enom\`enes critiques, syst\`emes al\'eatoires, th\'eories
  de jauge, Part II (Les Houches, 1984)}, pages 725--893. North-Holland,
  Amsterdam, 1986.

\bibitem[FW69]{FW69}
M.~E.~Fisher and B.~Widom. Decay of correlations in linear systems.
{\em J.~Chem.~Phys.}, 50: 3756--3772, 1969.

\bibitem[LF95]{LF95}
S.~Lai and M.~E. Fisher.
\newblock The universal repulsive-core singularity and {Y}ang-{L}ee edge
  criticality.
\newblock {\em J. Chem. Phys.}, 103:8144--8155, 1995.


\bibitem[Mil91]{Mil91}
J.~D. Miller.
\newblock Exact pair correlation function of a randomly branched polymer.
\newblock {\em Europhys. Lett.}, 16:623--628, 1991.

\bibitem[PF99]{PF99}
Y.~Park and M.~E. Fisher.
\newblock Identity of the universal repulsive-core singularity with
  {Y}ang-{L}ee edge criticality.
\newblock {\em Phys. Rev. E}, 60:6323--6328, 1999.
\newblock arXiv:cond-mat/9907429.

\bibitem[PS81]{PS81}
G.~Parisi and N.~Sourlas.
\newblock Critical behavior of branched polymers and the {L}ee-{Y}ang edge
  singularity.
\newblock {\em Phys. Rev. Lett.}, 46:871--874, 1981.

\bibitem[Sha83]{Sha83}
Y.~Shapir.
\newblock Supersymmetric dimer {H}amiltonian for lattice branched polymers.
\newblock {\em Phys. Rev. A}, 28:1893--1895, 1983.

\bibitem[Sha85]{Sha85}
Y.~Shapir.
\newblock Supersymmetric statistical models on the lattice.
\newblock {\em Physica D}, 15:129--137, 1985.

\bibitem[Sla99]{Sla99}
Gordon Slade.
\newblock Lattice trees, percolation and super-{B}rownian motion.
\newblock In {\em Perplexing problems in probability}, pages 35--51.
  Birkh\"auser Boston, Boston, MA, 1999.

\bibitem[SZK53]{SZK53}
Z.~W.~Salzburg, R.~W.~Zwanzig, and J.~G.~Kirkwood. Molecular distribution functions in a one-dimensional fluid. {\em J.~Chem.~Phys.}, 21:1098--1107, 1953.

\bibitem[Ton36]{Ton36}
L.~Tonks.  The complete equation of state of one, two, and three-dimensional gases of hard elastic spheres.  {\em Phys. Rev.}, 50:955--963, 1936.

\bibitem[WR70]{WR70}
B. Widom and J.~S. Rowlinson. 
\newblock New model for the study of liquid-vapor phase transition.
\newblock {\em Jour. Chem. Phys.}, 21:1670--1684, 1970.

\end{thebibliography}
\end{document}